# Analysis of Teacher's Management Policy: A Case Study of South Bangka, Indonesia

Raditya Bayu Rahadian [1(a)]

[1] Department of Education Management, Yogyakarta State University, Indonesia

[a] radityabayu.2020@student.uny.ac.id

## A. Introduction

It cannot be denied that teacher is the main factor to improve the quality of national education through their direct role in the learning process. In the Republic of Indonesia Law Number 20 of 2003 concerning the National Education System Article 39 it is stated that teachers as educators are professionals whose job is to plan and implement the learning process, assess learning outcomes, conduct mentoring and training, and conduct research and community service, especially for educators at universities. Furthermore, according to RI Law Number 14 of 2005 concerning Teachers and Lecturers it is stated that teachers are professional educators with the main task of educating, teaching, guiding, directing, training, assessing, and evaluating students in early childhood education through formal education, basic education, and secondary education.

In fact, it is explained in more detail regarding educator standards in RI Government Regulation Number 57 of 2021 concerning National Education Standards Article 20 paragraph (1) that educator standards are the minimum criteria for competence and qualifications possessed by educators to carry out their duties and functions as role models, learning designers, facilitator, and, student motivator. So deep and broad is the role of the teacher as the spearhead in realizing and actualizing learning, as well as living the ideals and ideas of an ideal education. Such a formidable task cannot be done haphazardly. Adequate provision of teacher capabilities is a very urgent matter for the government to carry out if it wants teachers to be able to carry out their duties and functions correctly according to what has been mandated by law.

There are at least two main problems faced by the government related to the teaching profession, namely regarding the quantity and quality of teachers. Not to mention if it is associated with the management of teachers with the regional autonomy system which has given birth to the domination of local governments in managing teachers. This resulted in teacher mobility being greatly hampered. The negative implications of hampered teacher mobility give rise to uncomfortable work psychology and the recruitment of teachers who are not yet professional is far from ideal.

Other classic problems that hit the teaching profession in the context of its management include; 1) minimal hiring of new teachers, 2) lengthy bureaucratic lines in force, c) difficult teacher data information system, d) recruitment and promotion of closed teachers, e) lack of strong commitment regarding hiring honorary teachers in general, f) distribution of teachers has not received serious attention in improving the quality of education, and g) the level of teacher competence is still low (Musfah, 2018).

Problems with teachers in the regions are still related to quantity and quality. Regarding the number of teachers,

Rahadian, R.B.                                                                                                  PAPERNIA Vol. 1, No. 1, February 2023the problems that are still encountered in South Bangka Regency include the unfulfillment of the ideal teacher-student ratio, the unequal teacher-student ratio in remote areas, and the low level of teacher equity among regions. With regard to teacher quality, the problems that still occur include low teacher competency, low teacher education levels, low number of teachers who have obtained educator certificates, and inequality in the distribution of teacher quality between regions (Dindikbud, 2020). The following data describes the condition of teachers in South Bangka Regency:

**Table 1. Teachers Condition in South Bangka**

| Area | Number of Students | Total Teachers | Teacher-Student Ratio | Bachelor Degree | Certified Teachers | Teacher Competence Test |
|---|---|---|---|---|---|---|
| Toboali | 12,774 | 647 | 1:20 | 97 % | 65% | 52 |
| Air Bara | 6,812 | 319 | 1:21 | 94 % | 43% | 50 |
| Payung | 3,471 | 127 | 1:27 | 94 % | 71% | 49 |
| Simpang Rimba | 3,967 | 152 | 1:26 | 92 % | 32% | 48 |
| Pulau Besar | 1,473 | 55 | 1:27 | 91 % | 78% | 40 |
| Tukak Sadai | 1963 | 68 | 1:29 | 81 % | 56% | 51 |
| Lepar Pongok | 1,361 | 64 | 1:21 | 84 % | 56% | 38 |
| Kep. Pongok | 774 | 35 | 1:22 | 86 % | 43% | 49 |
| Amount | 32,595 | 1,467 | 1:24 | 90% | 55% | 47 |

(Source: Educational Profil 2020 of South Bangka, Indonesia)

Table 1 shows that the average teacher-student ratio in South Bangka Regency is generally still reasonable, namely 1:24, meaning that a teacher supports 24 students. However, if we look at the sub-district data, we will find quite large discrepancies between sub-districts, namely between 1:20 to 1:29. There are sub-districts where the teacher-student ratio is quite high, reaching 1:29. This shows that the policy of equalizing teachers in South Bangka Regency is still problematic.

The qualification level of teacher education in South Bangka Regency is very good. The percentage of teachers who have a minimum education qualification of a Bachelor's degree has reached 90%. Most of the teachers have met the required educational qualification standards. This should show that the teacher has adequate competence in managing effective learning. However, the achievement of good educational qualifications is not directly proportional to the competence of teachers in implementing learning. This is indicated by the average result of the Teacher Competency Test (UKG) which is only 47. This condition shows the teacher's weak capability in implementing knowledge and experience in the learning process. If this condition continues, it will be the students who will be disadvantaged and will ultimately reduce the quality of regional education achievements. Policies taken in order to improve the quality of teachers are still low.

It makes sense that data presented in Table 1 can be concluded that the management of teachers is not caused by an ideal number of teachers and educational qualifications owned by teachers, but there are other problems that affect the quality of education in South Bangka Regency. Based on the results of a review of the program documents of the South Bangka Regency Education and Culture Office, there were several teacher quality problems that were the cause, including a mismatch between teacher qualifications and prospective teacher qualifications, formations that did not match needs, teacher transfers that were not based on teacher needs and capabilities, preparation inadequate teacher certification, and uneven opportunities for teachers to obtain professional competence development.

Broadly speaking, the division of regions in South Bangka Regency consists of urban, semi-urban, and remote/island areas. There are many gaps between the three regions, for example in cities there are many core teachers and outstanding teachers, in semi-urban areas, there are still many teachers who do not have a Bachelor's degree qualification, and in remote/island areas there are still many non-permanent (GTT) and honorary teachers. those who teach are not in accordance with their educational background and do not even meet the minimum qualifications of a Bachelor's degree.

Equal distribution of opportunities for teachers to develop their professional competence is a serious challenge for local governments. Limited resources and difficult geographical conditions, especially in remote areas and islands, mean that not all teachers can enjoy this opportunity. Welfare and access to information as well as opportunities to increase competence are specific constraints experienced by every teacher in the region. Therefore, equal opportunity for teachers, redistribution of teachers and the welfare of teachers in remote areas/islands need serious attention from the government and other stakeholders.

**B. MATERIAL AND METHOD**

This study uses a qualitative case study method based on the post-positivism philosophy to examine natural subject conditions, in which the researcher is the key instrument. Data collection techniques were carried out by means of triangulation (a combination of observations, interviews, and documentation). Data analysis is inductive (Sugiyono, 2019: p.25).

This research was conducted in South Bangka, Indonesia from January to June 2022, involving respondents including the Regional Secretary, Head of the Education Office, Principals, and other government officials who handle education affairs in South Bangka, Indonesia.

This study aims to determine the teacher management policies that have been implemented in South Bangka, Indonesia. These results are then reviewed and analyzed based on theory and scientific literature to obtain a balance between the advantages and disadvantages of the policies that have been carried out as suggestions for more in-depth research in the future.

**C. RESULTS**

The South Bangka Government is trying to make policies aimed at improving the quality of teachers and regional education. Several policies have been implemented related to the management of teacher resources in addressing issues of planning, placement, distribution, and teacher development, namely:

**1. Teacher Needs Analysis**

Analysis of teacher needs is used to develop personnel requirements/data elements, which must be linked to strategic organizational planning, budgeting processes, and all recurring recruitment needs, training recruitment, and planning activities (Mukminin, Habibi, Prasojo, & Yuliana, 2019: p.116) . The analysis of teacher needs was carried out by the South Bangka Regency

27

Government, in this case, the South Bangka Regency Education and Culture Office, which is still struggling with administratively adjusting teacher data so that it is accepted into the Basic Education Data system (Dapodik). Therefore the real problems that occur in the field are not really resolved, because the focus of resolution is only on the administrative level. If the data has been received at Dapodik, no follow-up is carried out to unravel real problems in the field. Whereas problems in the field related to excess teaching hours, shortage of teaching hours, shortage of teachers in a school, equal distribution of quality and capability of teachers between regions, transparent teacher recruitment process, and as needed must be resolved in real terms, not only administratively.

Recommendations related to the system of recruitment and equalization of teachers and improving the quality of teachers are as follows: a) the applicable policy regarding hiring (recruitment) of honorary teachers conforms to the procedures for recruitment of civil servant teachers, b) the system of transferring teachers in districts is carried out with a needs analysis and not based on likes and dislike as well as political orders, c) the development of a teacher quality improvement program that takes into account regional equity and does not focus on certain schools (not only in cities or only high-achieving schools).

**2. Participatory Management in the Teacher Redistribution Program**

Teacher distribution policies through participatory management can be understood as empowering teachers to participate in making decisions related to efforts to distribute teachers in various schools. The provisions for a minimum teaching load of 24 hours face-to-face for teachers to obtain regional allowances and certification allowances can be used by local governments to involve teachers' participation in making related decisions.

Participatory management has many positive values, namely: 1) the opportunity to participate in formulating policies, is an important factor for increasing teacher morale, 2) involvement in decision-making has a positive relationship with teacher satisfaction, 3) teachers prefer to be involved in decision making, and 4) teachers who are involved in making decisions in fact make a good contribution in the decision-making process (Winoto, 2020: 161) .

Teachers in cities that are short on teaching hours due to a surplus of teachers can be involved in determining the redistribution of teachers to other areas. Fulfilling the teacher's workload of at least 24 (twenty-four) face-to-face hours in 1 (one) week forces teachers to be able to fulfill their duties. If the teacher cannot fulfill it, allowances related to the teacher's functional position cannot be paid. The willingness of teachers to choose schools in rural areas where there is a relative shortage of teachers to fulfill the obligation to fulfill 24 hours of lessons/week will minimize the risk of rejection which will have a negative impact on the psychology of the teacher which can impact students.

**3. Multigrade Learning**

Multi-grade learning is the policy chosen to address teacher shortages in remote/archipelagic areas. In a multigrade class, one teacher will provide learning to two study groups (groups) or more that come from the same or different levels. For example grades 1 and 2 are taught by the same teacher. Multi-grade learning pays attention to the class level being taught so that it does not make it difficult for teachers and students to implement it. This means that the level of similarity becomes a reference in multigrade learning. Lower grades which include grades 1, 2, and 3 can be combined to be taught by one teacher. Likewise, grades 4, 5, and 6 can be combined to be taught by one other teacher.

Multi-grade learning conducted in remote/archipelago areas is forced to have no other choice due to the limited number of available teachers. The main consideration is equality of opportunity to obtain educational services for the entire community. Even so, there are several things that need to be done by local governments to improve the quality of learning services for schools implementing multigrade, including 1) giving priority to teachers conducting multigrade to get opportunities to improve their capabilities through various guidance and training, and 2) guarantee the feasibility of teacher data in Dapodik to avoid data errors that can harm teachers.

The local government can also attract local residents who are committed to being educated at the expense of the government to then be assigned to the area. A clear commitment and good oversight will ensure that this program produces quality teachers

**4. Roving Teacher**

Itinerant teachers are teachers who are given the task of teaching and/or being involved in developing the quality of education in other schools outside of the school where they serve for a certain period of time without being transferred from their original school. This policy is a system for periodically transferring teaching assignments to teachers in districts/cities. In general, the objectives of implementing mobile teachers, in general, have the objectives of 1) equalizing the quality of learning and learning in schools that need teachers, 2) equalizing the ability and professional quality of teachers, 3) equalizing the workload of teachers, 4) lightening the workload of teachers who too heavy in remote areas, and 5) improve the ability/performance of teachers evenly in all schools.

The need for teacher competency development with a limited budget can take advantage of this mobile teacher policy. The benefits obtained are: 1) making it easy for teachers to study in order to improve their competence and professional abilities as well as improve their Bachelor's qualifications without leaving their teaching duties, 2) providing policy space for teachers to meet the total teaching hours load of 24 lesson hours/week, 3) increasing the efficiency of education management in the district, 4) for the principal teachers who are involved as mobile teachers, indirectly they will continuously develop their performance and competence in the pedagogic, professional, social, and personality fields, and 5) for teachers who are certified educators to have the opportunity to show their skills and continuously develop their professional abilities through their participation in the itinerant teacher policy.

Another benefit of the mobile teacher policy is that schools with a shortage of teachers feel helped by the presence of several other teachers. Meanwhile, teachers who have less than 24 hours/week of lessons can take advantage of this program to act as mobile teachers so they don't experience a shortage of teaching hours anymore. Meanwhile, another benefit of this mobile teacher is that students feel motivated and not bored. This is due to the ever-changing learning atmosphere and the teachers who teach are always changing so that various different experiences will be obtained by students.

### 5. In-service Training

In-service training is an effective method for positively increasing the knowledge, skills, and confidence of teachers. This is the process teachers use to continue their education after they have received certification in teaching and working professionally (Ary, Jacobs, & Sorensen, 2010: p.164). In-service training is a term that describes a series of activities and requirements relating to professional development, organized to improve the performance of all personnel (teachers) who have been assigned and hold positions in schools so that they are able to implement programs or innovations.

In-service training aims to increase and improve the knowledge, skills, and experience of teachers in carrying out their duties and obligations. The in-service training program can include various activities such as holding workshops, lectures, holding teacher-teacher meetings to exchange experiences, seminars, visits to schools outside the region and special preparations, and attending training related to the main task. . In-service training activities are one of the key factors in the professional development of teachers and provide an increase in knowledge through the active role of the teacher.

The implementation of in-service training in South Bangka Regency in terms of quantity is still rarely implemented because in-service training activities require a large enough budget, which causes a small opportunity for teachers to participate. Empirically, in-service training activities benefit teachers in improving skills, competencies, and managing training more effectively. One alternative that can be done is to take advantage of the available budget slots in the School Operational Assistance (BOS) funds according to technical guidelines that can be allocated to increase teacher professional competence. The local government through the South Bangka Regency Education and Culture Office must provide adequate socialization and budget simulations so that schools can implement them properly.

### D. Discussion

Each policy must have its own advantages and disadvantages, however ideally a policy. This is of course related to the many components involved in education. Based on the several policies that have been described, the advantages and disadvantages of each policy can be analyzed, as follows:

#### 1. Teacher Needs Analysis

Strengths: 1) being able to find out the number of teacher needs needed, 2) obtaining alternative substitutes and solutions to overcome teacher shortages.

Disadvantages: 1) cannot be realized directly because they have to wait for the budgeting process, 2) requires accurate data to carry out the analysis process, and 3) the implementation process takes a long time because it is necessary to conduct field surveys first.

#### 2. Participatory Management in the Teacher Redistribution Program

Strengths: 1) involving teachers in decision-making: 2) increasing teacher willingness and responsibility in its implementation.

Disadvantages: 1) sharp debates often lead to conflicts that affect the work atmosphere, 2) not all teachers are open-minded to reach problem-solving.

#### 3. Multigrade Learning

Strengths: 1) can overcome the shortage of teachers and classrooms, 2) can save teaching time, and 3) teachers are more challenged to be more creative in the learning process.

Disadvantages: 1) teachers' abilities are not evenly distributed, 2) requires more effective time management, 3) requires greater teacher creativity, innovation, and ability, 4) the level of achievement of the curriculum is often incomplete, 5) lack of respect for individual differences in participants taught students.

#### 4. Roving Teacher

Strengths: 1) can be used to overcome the shortage of teachers, 2) can produce qualified teachers, 3) helps fulfill the obligation to teach 24 hours/week, 4) expands the knowledge and experience of teachers, 5) teachers feel challenged to develop their competence, 6) the quality of education can be distributed to target schools, 7) the learning atmosphere is fresher.

Disadvantages: 1) it is geographically difficult due to the spread of schools in remote/island areas, 2) requires careful consolidation regarding teacher readiness, and 3) requires stronger energy and mind because they have to move around in teaching.

#### 5. In-service Training

Strengths: 1) improve teacher capabilities, 2) can be managed more effectively.

Disadvantages: 1) requires a relatively large budget, 2) it is possible that there are many vacancies in-class hours because the teacher often leaves them to attend training activities, 3) the material received is not continuous and often makes it boring, 4) the opportunity to participate in training activities is relatively small because the number of training activities few while teachers who will follow very much.

### E. Conclussion

The various policies taken by the South Bangka Regency Government, in this case, the South Bangka Regency Education and Culture Office, seek to improve the quality of teachers and learning in the midst of conditions and limited resources. Every policy taken must have considered the impact it will have, both positive and negative. Whatever education policy is taken, in essence, it must be in favor of the general public and have a greater positive impact on improving the quality of regional education services. Careful planning and fulfillment of the stages in determining educational policies must be carried out strictly so that they can be used to control the desired impact. In addition, policy sustainability must be considered and not only based on momentary trends.